**Title: Investigation of the Relationship between Geomagnetic Activity and Solar Wind Parameters Based on A Novel Neural Network (Potential Learning)**

Author #1: Ryozo Kitajima

Department of Engineering, Tokyo Polytechnic University, 1583 Iiyama, Atsugi City, Kanagawa Prefecture 243-0297, Japan.

r.kitajima@eng.t-kougei.ac.jp

Author #2: Motoharu Nowada

Shandong Provincial Key Laboratory of Optical Astronomy and Solar-Terrestrial Environment, Institute of Space Sciences, Shandong University, 180 Wen-Hua West Road, Weihai City, Shandong Province, 264209, People's Republic of China.

moto.nowada@sdu.edu.cn

Author #3: Ryotaro Kamimura

IT Education Center, Tokai University, 4-1-1 Kitakaname, Hiratsuka City, Kanagawa Prefecture 259-1292, Japan.

ryo@keyaki.cc.u-tokai.ac.jp

**Corresponding authors: Ryozo Kitajima and Motoharu Nowada**

# Abstract

Predicting geomagnetic conditions based on in-situ solar wind observations allows us to evade disasters caused by large electromagnetic disturbances originating from the Sun to save lives and protect economic activity. In this study, we aimed to examine the relationship between the $K_p$ index, representing global magnetospheric activity level, and solar wind conditions using an interpretable neural network known as potential learning (PL). Data analyses based on neural networks are difficult to interpret; however, PL learns by focusing on the "potentiality of input neurons" and can identify which inputs are significantly utilized by the network. Using the full advantage of PL, we extracted the influential solar wind parameters that disturb the magnetosphere under southward Interplanetary magnetic field (IMF) conditions. The input parameters of PL were the three components of the IMF ($B_x$, $B_y$, $-B_z(B_s)$), solar wind flow speed ($V_x$), and proton number density ($N_p$) in geocentric solar ecliptic (GSE) coordinates obtained from the OMNI solar wind database between 1998 and 2019. Furthermore, we classified these input parameters into two groups (targets), depending on the $K_p$ level: $K_p$ = 6- to 9 (positive target) and $K_p$ = 0 to 1+ (negative target). Negative target samples were randomly selected to ensure that numbers of positive and negative targets were equal. The PL results revealed that solar wind flow speed is an influential parameter for increasing $K_p$ under southward IMF conditions, which was in good agreement with previous reports on the statistical relationship between the $K_p$ index and solar wind velocity, and the $K_p$ formulation based on the IMF and solar wind plasma parameters. Based on this new neural network, we aim to construct a more correct and parameter-dependent space weather forecasting model.

# 1. Introduction

The terrestrial magnetosphere protects life from the harmful radiation effects associated with the high-speed plasma streams (solar wind) and is constantly undergoing dynamic changes due to interactions with solar wind and the interplanetary magnetic field (IMF) originating from the Sun, effective (e.g., Black 1967; Glassmeier et al. 2009; Glassmeier and Vogt 2010). Drastic changes from quiet to active geomagnetic conditions start from a violation of the "frozen-in-condition" of the geomagnetic field caused by reconnecting the geomagnetic field with solar wind field lines, known as magnetic reconnection. Substorms, magnetic storms and auroral signatures are phenomena observed in the magnetosphere that occur due to reconnection-associated transfers of solar wind energy into the magnetosphere, and the resultant magnetospheric activity is of a high level.

The $K$ index, defined as the value representing the level of geomagnetic disturbances driven by the solar wind based on the perturbations in the Earth's magnetic field, was used to determine geomagnetic conditions. Moreover, it defines geomagnetic disturbances using an integer in the range 0–9 with 1 being calm and 5 or more indicating a geomagnetic storm. This index was first introduced by Bartels (1939) and is derived from the maximum fluctuations of horizontal components observed on a magnetometer with a temporal resolution of 3 h. Today, the $K_p$ index, derived from the weighted average of the $K$ indices of 13 geomagnetic observatories around the world (Bartels 1949), is understood to be the most representative proxy parameter for measuring energy input from solar wind to Earth and the resultant geomagnetic activity. Other examples of indices representing geomagnetic activity use the current intensity associated with aurora ($AL$ and $AU$ indices) and the strength of looped currents (ring current) flowing around the magnetic equator region built up by magnetic storms ($D_{st}$ index). Since their inception, $K_p$ values have been used for important and reliable index to representations of global geomagnetic activity. However, the time resolution of $K_p$ (2.5 h) is lower than those of the other geomagnetic indices, such as $AL$, $AU$, PC (Polar Cap), and $D_{st}$, which have time resolutions ranging from 1 min. to 1 h (see Rangarajan 1987).

Derivations of these indices to show global or specific region geomagnetic activity based on solar wind parameters have been conducted. In particular, $K_p$ has been derived from solar wind parameters at Lagrange point 1 (L1) obtained by satellites (e.g., Wing et al. 2005; Wintoft et al. 2017; Zhelavskaya et al. 2019; Shprits et al. 2019). Newell et al.

(2008) formulated the $K_p$ index based on the IMF and solar wind–magnetosphere coupling functions, which are equations that quantitatively evaluate the amount of solar wind energy inputs to the magnetosphere based on IMF and solar wind plasma parameters (Newell et al. 2007). The equations are as follows:

$$K_p = 0.05 + 2.244 \times 10^{-4} \left(\frac{d\Phi_{MP}}{dt}\right) + 2.844 \times 10^{-6} N_p^{\frac{1}{2}} V_{sw}^2 \qquad (1)$$

$$\frac{d\Phi_{MP}}{dt} = V_{sw}^{\frac{4}{3}} B_t^{\frac{2}{3}} \sin^{\frac{8}{3}}\left(\frac{\theta_{clock}}{2}\right) \qquad (2).$$

According to Eq.(1), $K_p$ can be represented by the solar wind proton number density ($N_P$); velocity ($V_{sw}$); IMF clock angle, defined as the angle between the IMF-$B_y$ and -$B_z$ components ($\theta_{clock}$ = arctan(IMF-$B_y$/IMF-$B_z$)); and IMF intensity ($B_t$), which is included in Newell's solar wind–magnetosphere coupling function ($d\Phi_{MP}$/dt), as calculated using Eq.(2). Eqs. (1) and (2) show that the solar wind velocity is closely correlated with the $K_p$ index, as advocated by studies by Snyder et al. (1963) and Elliott et al. (2013). Nevertheless, it is difficult to determine the geomagnetic disturbance level based on solar wind conditions because of the complicated relationships between geomagnetic activity, IMF, and solar wind plasma.

Recently, machine learning (or deep learning) approaches have been used to predict $K_p$.

The artificial neural network (NN) is one of the most popular algorithms for $K_p$, $D_{st}$ and PC forecasting (e.g., Nagai 1994; Costello 1998). Later, Boberg et al. (2000) and Wing et al. (2005) developed a prediction model based on NN using with IMF and solar wind plasma as input parameters. Boberg et al. (2000) sequentially built a multi-layer feed-forward network using IMF-$B_z$ component, solar wind plasma density ($N_p$), and velocity ($V_{sw}$) as the input parameters, and evaluated the developed algorithm in terms of "training", "validation", and "test" based on the correlation and root-mean-square error (RMSE). Furthermore, an NN was developed by Bala and Reiff (2012) to forecast three indices: $K_p$, $D_{st}$, and $AE$ (as defined by $AU - AL$). They obtained and compared several forecasting patterns of the $K_p$ index with various solar wind input parameters and found significant differences in the RMSE and correlation between the obtained models. They also evaluated the prediction time for forecasting performance and concluded that RMSE tends to become larger as $K_p$ prediction time increases.

Following these NNs, Ji et al. (2013) introduced a support vector machine (SVM) to build a $K_p$ forecasting model and evaluated the forecasting results from SVM by comparing the $K_p$ prediction results with those from an NN. They constructed a forecasting model under high magnetic activity conditions. Tan et al. (2018) constructed and evaluated a $K_p$ forecasting model using the solar energy input function (a coupling function) and the associated viscous term as inputs (Newell et al. 2008). Their models can also consider the $K_p$ forecasting error and were built based on long short-term memory (LSTM), which was developed from recurrent NNs (RNNs) (Hochreiter and Schmidhuber 1997).

In this study, we developed an extraction algorithm for solar wind parameters, significantly affect geomagnetic disturbances, with the help of the $K_p$ classification model based on potential learning (PL). PL has been used to conduct analyses where high model performance and high interpretability are required. For example, in a study that applied PL to supermarket data (ID-POS) by Kitajima et al.(2016a), a model was developed that used the "consumer's purchase behavior in the past three months" as an input parameter to determine the "customer's probability to visit the store in two months in the future". They determined that the model based on PL performed better than the conventional method and succeeded to extract an important variable. In addition, PL has been applied to data in various fields, such as Tweet data at the time of a disaster (Kitajima et al. 2016b) and data on president messages of the companies (Kitajima et al. 2019). Since PL has been used for data analysis in various fields, we aim to identify the most significant parameters that disturb the magnetosphere based on PL. Furthermore, we will run several PLs by changing the parameters and evaluate the performance of the application of PL to space physics data. We will also compare the results obtained from PL with those from another algorithm, multi-layer perceptron (MLP), and discuss the difference between the two algorithms.

This paper is organized as follows. Section 2 presents the data used, and methodology in this study. The evaluation of the performance of PL, the results obtained based on PL and the differences between PL and MLP are shown in section 3. In section 4, finally, we present the discussion and our conclusions of this study. In Appendix, we describe the details of the PL structure.

## 2. Data and Methodology

### 2.1 Database compiling

In this study, we used the three components of the interplanetary magnetic field (IMF;

Bx, By, Bz), and solar wind plasma parameters, such as solar wind velocity and ion number density, in geocentric solar ecliptic (GSE) coordinates and the global geomagnetic activity index ($K_p$ index) from January 1 1998 to December 31 2019, as input parameters for PL. Detailed information on the parameters of the solar wind and geomagnetic activity index is summarized in Table 1. The solar wind parameters with temporal resolution of 1 min. of the OMNI database and the $K_p$ index with a time resolution of 3 h were utilizes, respectively. We calculated the 3 h average of the solar wind data to give these parameters the same temporal resolution as $K_p$. If the parameter had a data gap larger than 40%, the averages were not computed. To further exclude the observation data in the magnetosphere from the database, we established a threshold where the satellite GSE-X component (sun-earthward) was larger than the nominal nose point (~15 $R_E$) of the model bow shock, proposed by Farris and Russell, (1994), that is, the database used completely comprised the observation values in interplanetary space.

In this study, we considered only magnetospheric activity under the southward (negative) IMF-$B_z$ case and excluded the northward (positive) IMF-$B_z$ component as this NN input parameter, identified with "$B_s$" in Table 1. There were two main reasons for this criterion for the IMF-$B_z$ component. First, we considered that geomagnetic conditions are favorable to be disturbed because high occurrences of magnetic reconnection can be expected in dayside magnetosphere. Second, PL learns by focusing on the highest variance of the parameters (see Eqs. (3) and (4) in section 2.3) and extracts the focused parameter as the most significant factor driving the magnetospheric disturbances. Therefore, we excluded cases of the IMF-$B_z$ component highly fluctuating between positive and negative around 0 nT.

Before inputting the solar wind conditions to the PL, we classified the $K_p$ values into two groups (targets) of "positive" and "negative" targets. $K_p$ index with values from 6- to 9, and the associated solar wind data were labeled as "positive target (group)". Whereas $K_p$ values ranging from 0 to 1+ and the associated solar wind parameters were labelled as "negative target (group)". The total number of compiled (averaged) data points was 27,168 with the positive (negative) target number being 793 (26,375). To equalize the number of data between the positive and negative targets, we randomly chose and extracted 793 points out of the 26,375 negative target data points. Finally, we analyzed 1,586 positive and negative data points.

**2-2 Methodology of database analysis**

By adopting a new NN (PL) to take the 3 h average solar wind parameters as "input parameters" and classify whether or not the associated $K_p$ index belongs to "positive" or "negative" targets, we investigated the relationship between geomagnetic activity levels and solar wind parameters.

Recently, NNs have been adopted to analyze databases with complicated structures in space plasma physics. In general, NNs have frequently been used to build forecasting models of geomagnetic indices; however, it is difficult to interpret which solar wind parameters are the most important in disturbing the magnetosphere. In this study, we applied a new NN theory, PL, which was developed based on two NNs; selective potentiality maximization, proposed by Kamimura and Kitajima (2015), and self-organizing selective potentiality learning (Kamimura 2015).

In this study, we trained the PL by setting the number of neurons (see the details on the manner to train PL are described in section 2.3), as listed in Table 2. The number of hidden neurons was automatically determined by the software of "SOM Toolbox v2.1" which was developed by Vatanen et al. (2015).

In the knowledge utilization step, hyperbolic tangent and softmax functions were used for the activation functions of the hidden and output neurons, respectively. We searched for the most suitable value by varying the value of parameter "r" from 1 to 10 with a step of 1. A total of 1,110 (70%) of the 1,586 samples were used for training. Half of the remaining 238 samples (15%) were utilized to prevent training from overfitting (early stopping) and the other half (15%) were used for testing. We maintained these allocation rates during these PL runs. In this study, we made 10 different models in a random choice manner, as shown in Figure 2. We evaluated the performance of each model by calculating the average values of the 10 models.

**2.3 Details of the potential learning (PL)**

PL consists of two steps: knowledge accumulation, based on self-organizing maps (SOM), the concept of which is shown in Figure 1(a); and knowledge utilization, originating from multi-layer perceptron (MLP), the details of which are shown in Figure 1(b). During knowledge accumulation, the potentiality of the input neuron is calculated and knowledge is acquired (training). Here, we define "potentiality" as ability which can response to various conditions of neuron. In case of "Neuron with high potentiality", it indicates the neuron which can play an important role in training. In general, NNs are referred to as "black box," but, in the PL, we can interpret which input parameters are important by interpreting the potentiality after training. If assigning the number $k$ ($k = 1,2,...,K$) to the input neuron, we can derive the potentiality of the $k^{th}$ input neuron ($\Phi_k^r$) between 0 and 1, using the following equation:

$$\Phi_k^r = \left( \frac{V_k}{\max_{k=1,...,K} V_k} \right)^r \qquad (3).$$

where $V_k$ is the variance of the $k^{th}$ input neuron, which is computed based on "weight" ($w_{j,k}$) connected to the $k^{th}$ input neuron from $j^{th}$ ($j = 1,2,...,J$) output neuron and $r$ is the parameter that controls the potentiality calculated using the algorithm. The larger the "r" value becomes, the input neuron with larger variance can have larger potentiality.

After the potentiality was calculated, PL was trained based on self-organizing maps (SOM), in which the potentiality was used to calculate the distance ($d_j$) between the input neuron (the input from the $k^{th}$ input neuron is denoted by $x_k$) and the $j^{th}$ output neuron with the following formula:

$$d_j = \sqrt{\sum_{k=1}^{K} \emptyset_k^r (x_k - w_{j,k})^2} \qquad (4).$$

Eq. (4) means that the "distance," weighted by the potentiality of the input neuron, was used in the training process. The logics for the other training were the same as those for the SOM. Through the Knowledge accumulation step, PL starts to conduct the training at the step of Knowledge utilization, based on MLP. In this step, the weight obtained in the knowledge accumulation step was multiplied by the potentiality and set as the initial weight between the input and hidden layers for learning. In general, the results of the training based on MLP depend on the initial weights. However, PL is expected to provide more precise training based on the knowledge obtained from the input parameters (data).

## 3. Results

**3.1 Evaluation of model performances**

To evaluate the PL 10 models, we calculated the values of four measures (accuracy, precision, recall and F-measure) with changing value of parameter "r" from 1 to 10. Table 3 shows the calculation results of the four measures, indicating the extent to which the model successfully predicted the test data. When "r" was 6, the value of "accuracy" was the highest. The main purpose of creating the 10 models was to extract of the variables that play essential roles in classifying the $K_p$ index into two targets: negative and positive. Therefore, we focused on the case with the highest accuracy value and thus applied the best model with r = 6 in this study.

We also compared the test results based on PL with those of MLP, a basic NN. As shown in Table 3, all values (accuracy, precision, recall and F-measure) in MLP were close to those of MLP. In particular, the difference in accuracy between PL and MLP was only 0.0063. MLP can be better for classifying $K_p$ into two targets than PL, if the main purpose is only the prediction of geomagnetic activity. However, PL actively selects the input values to be utilized for classification while MLP does not.

When r = 6, the values of three measures (accuracy, recall, and F-measure) of evaluating model performance in PL reached their maximum, but were slightly lower than those in MLP. Precision reached its maximum at r = 10. PL, however, has a strong advantage in extracting the most influential solar wind parameters that cause geomagnetic disturbances. Therefore, in this study, we applied the PL model with r = 6.

## 3.2 Extraction of significant solar wind parameters that cause magnetospheric disturbances

Figure 3 shows the result of PL for the input neurons at r = 6. PL extracted the solar wind velocity ($V_x$) as the parameter with the highest "input potentiality" (~1.0), suggesting that PL at r = 6 judged solar wind velocity to be the most significant parameter causing geomagnetic disturbances under the Bs (southward IMF) condition. The parameter with the next highest potentiality was the solar wind density ($N_p$) at 0.0431; however, this can almost be ignored when compared with the potentiality of the solar wind velocity.

Figure 4 shows the weights of the input and hidden layers used in the PL and MLP for comparison. The length of the bar shows the weight in PL, and the signs of the weight values are indicated with red (plus) and green (minus), respectively. The panel (a) in Figure 4 shows that MLP has various plus and minus values for weights at each input neuron (parameter), indicating that it is difficult to identify which input neuron (parameter) was used in the network. However, in the PL network with r = 6 (panel b), most of the weight was concentrated on the third variable (solar wind velocity); however, the fourth (ion number density) and fifth (southward IMF) variables also had some weight and were thus also used in the PL network. PL uses potentiality to set up the initial weight in the knowledge utilization step (see Figure 1b). Although three variables (third, fourth, and fifth) had high weight values, we judged the parameter with the highest weight value, the third variable (solar wind velocity), as having the most significant potentiality among them.

## 4. Summary and Discussion

We reported the results of benchmarks of the application of a new neural network (Potential Learning) for the prediction of geomagnetic activity, driven by solar wind, and the successful extraction of the most significant solar wind parameter in causing geomagnetic field disturbances. This study is the first attempt for applying the PL to the numerical data analyses in space plasma. We also used 22 years of OMNI solar wind data and $K_p$ indices as input neurons but only used the data when the IMF $B_z$ was southward. This was because geomagnetic activity is favorable to be disturbed by dayside magnetic reconnection under southward IMF-$B_z$ conditions (e.g. Dungey, 1961), and it is thus easier to extract the crucial solar wind parameter(s) that drive the geomagnetic disturbances.

We excluded the solar wind data under northward IMF conditions due to an inherent disadvantage of the current PL algorithm; PL identifies the largest variance value with the highest potentiality. Therefore, if data under northward IMF conditions were included in the database, the stable (non-excursive) but intensive southward IMF-$B_z$ component cannot be chosen as the solar wind parameter with the highest potentiality. Furthermore, the fluctuating IMF-$B_z$ around 0 nT may be chosen as the most significant parameter that cause magnetospheric disturbance. To avoid these cases, we utilized only solar wind data during the southward IMF intervals as input neuron. In future studies, we need to improve the PL algorithm, which applies an importance to the largest variance value for the highest potentiality.

Based on a large solar wind database, PL extracted the solar wind velocity as the parameter with the highest potentiality when r = 6 (see Figure 3), suggesting that solar wind speed ($V_x$) is an important parameter in disturbing geomagnetic conditions. Significant enhancements of the global geomagnetic activity level due to increases in the

$V_x$ component were reported by Snyder et al. (1963). More recently, Elliott et al. (2013) examined the relationship between the $K_p$ index and solar wind speed, separating into low and high solar wind number density and dynamic pressure cases and the presence/absence of solar wind disturbances, such as the interplanetary coronal mass ejection (ICME). Furthermore, Thomsen (2004) suggested that the large-scale convection electric field ($E_c$ = -$V_{sw}$ x $B_{geo}$), calculated using the solar wind velocity ($V_{sw}$) and geomagnetic field ($B_{geo}$), has a good correlation with the $K_p$ index. This quantitative relationship between solar wind velocity and the $K_p$ index, supported by the two velocity terms of "$V_{sw}^2$" and "$V_{sw}^{3/4}$" being comprised in a formulation proposed by Newell et al. (2008) (Eq. 1), suggests that solar wind velocity is the most important parameter in controlling $K_p$. Therefore, the most significant parameter extracted by PL (solar wind velocity) was determined to be the most significant parameter that causes disturbances to the Earth, being consistent with previous statistical observational results (Gholipour et al. 2004; Newell et al. 2008; Elliott et al. 2013, and references therein).

Comparing the MLP results with those of PL, the accuracies were not significantly different. However, PL could be applied to extract the most significant parameter leading to space weather disasters from solar wind. This benchmark for the application of PL to the space weather-related problem verified its effectiveness in predicting the solar wind driving geomagnetic activity and significant solar wind parameters that cause geomagnetic disturbances.

In this study, we ensured that PL can extract the most significant solar wind parameter which causes geomagnetic disturbances. Therefore, we can aim to construct a more correct and parameter-dependent space weather forecasting model based on PL.

## Declarations

### List of abbreviations

PL: Potential Learning; IMF: Interplanetary Magnetic Field; RMSE: Root-Mean-Square Error; NN: artificial Neural Network; MLP: Multi-Layer Perceptron; GSE coordinates: Geocentric Solar Ecliptic coordinates; SOM: Self-Organizing Maps

### Availability of data and materials

Solar wind OMNI data were obtained from the Coordinated Data Analysis Web

(https://cdaweb.sci.gsfc.nasa.gov/index.html/), provided by GSFC/NASA. $K_p$ index data were provided by the World Data Center for Geomagnetism, Kyoto (http://swdcdb.kugi.kyoto-u.ac.jp/).

**Competing interests**

The authors declare that they have no competing interest.

**Funding**

This study was supported by a grant from the National Natural Science Foundation of China (NSFC 42074194) (M.N.).

**Authors' contributions**

Motoharu Nowada conceived the research project. Ryozo Kitajima performed all data analyses, made all the figures, and tuned the PL codes. Motoharu Nowada and Ryozo Kitajima wrote the paper and edited the manuscript. Ryotaro Kamimura developed the main engine of the PL program and edited the draft.

**Acknowledgements**

We would like to thank Editage (www.editage.com) for English language editing.

**Authors' information**

[1]Department of Engineering, Tokyo Polytechnic University, 1583 Iiyama, Atsugi City, Kanagawa Prefecture 243-0297, Japan. [2] Shandong Provincial Key Laboratory of Optical Astronomy and Solar-Terrestrial Environment, Institute of Space Sciences, Shandong University, 180 Wen-Hua West Road, Weihai City, Shandong Province, 264209, People's Republic of China. [3]IT Education Center, Tokai University, 4-1-1 Kitakaname, Hiratsuka City, Kanagawa Prefecture 259-1292, Japan.

**Figure legends**

Figure 1. The concept of potential learning (PL). PL comprises two important steps: (a) knowledge accumulation and (b) knowledge utilization.

Figure 2. Block diagrams of details of 10 potential learning (PL) models. In each model, the data for training (Training data), the data to prevent training from overfitting (early stopping) (Validation data), and the data for testing the model (Testing data) are included. The percentages for the three kinds of data are 70%, 15%, and 15%, respectively.

Figure 3. Results of the application of PL at r = 6. The five OMNI solar wind parameters (IMF-$B_x$, IMF-$B_y$, $V_x$, $N_p$, and $B_s$) are chosen as the input data to PL. The horizontal and vertical axes give input potentiality and the numbers of input five solar wind parameters, respectively. The potentialities of the solar wind velocity and density are 1.0 and 0.0431, respectively.

Figure 4. Weights in the input – hidden layers in the networks of MLP (a) and PL (b). Horizontal and vertical axes give the number of five input variables (neurons) and number of neurons in hidden layer, respectively. The length of the bar shows the weight in PL, and the signs of the weight values are indicated with red (plus) and green (minus),

respectively.

**Table legends**

Table 1. Detailed information on the parameters in compiled database used in this neural network

Table 2. List of numbers of neurons in potential learning (PL)

Table 3. Summary of accuracy, precision, recall and F-measure values. Bold letters indicate their maxima

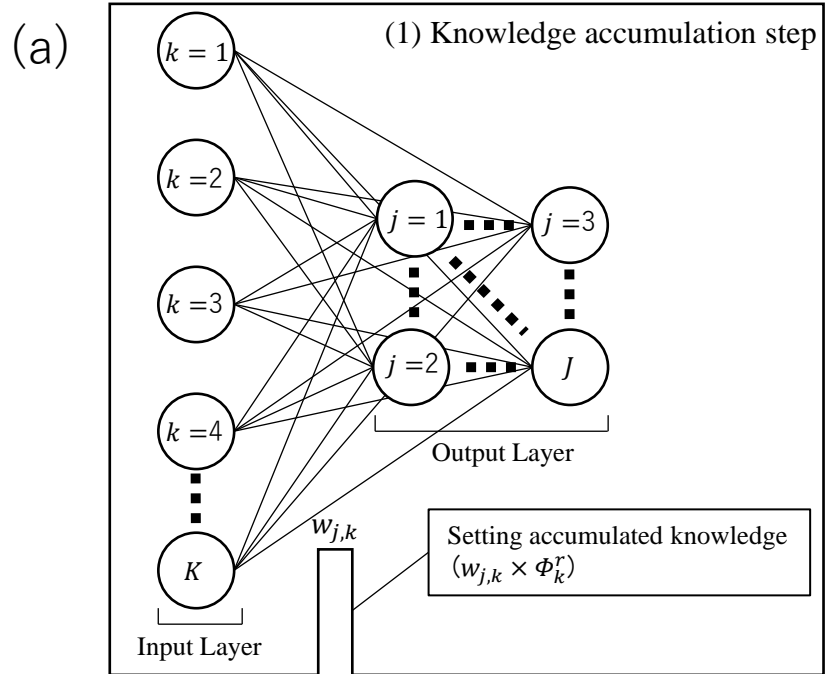
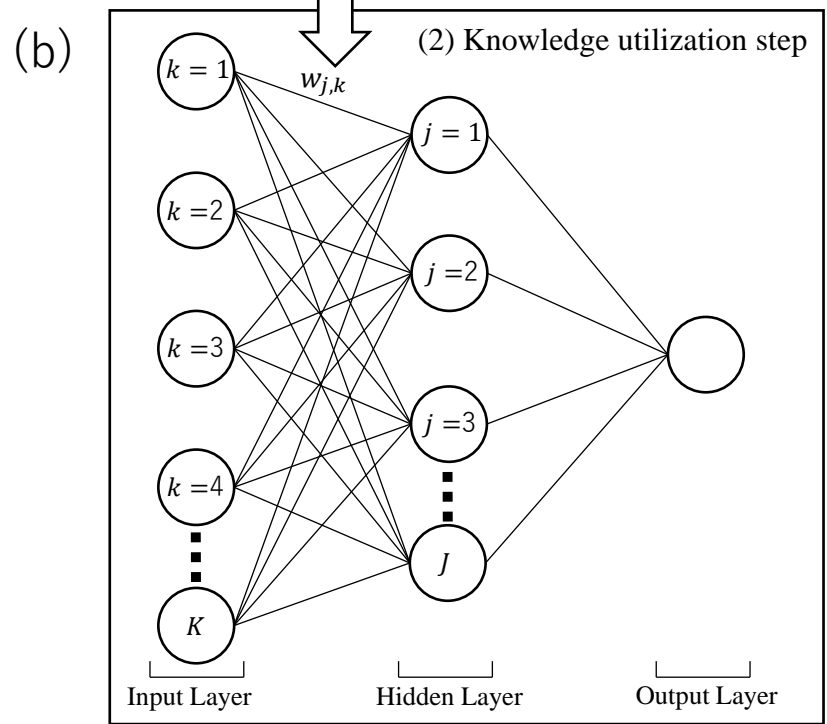

Figure 1

Figure 2

| No. | Purpose |
|---|---|
| 1 | Training data |
| 2 | Training data |
| 3 | Training data |
| 4 | Validation data |
| 5 | Validation data |
| ⋮ | Testing data |
| n | Testing data |

70%
15%
15%

Data pattern 1

| No. | Purpose |
|---|---|
| 1 | Testing data |
| 2 | Testing data |
| 3 | Validation data |
| 4 | Validation data |
| 5 | Training data |
| ⋮ | Training data |
| n | Training data |

15%
15%
70%

Data pattern 2

·········

| No. | Purpose |
|---|---|
| 1 | Validation data |
| 2 | Testing data |
| 3 | Training data |
| 4 | Validation data |
| 5 | Training data |
| ⋮ | Testing data |
| n | Training data |

15%
15%
70%

Data pattern 10

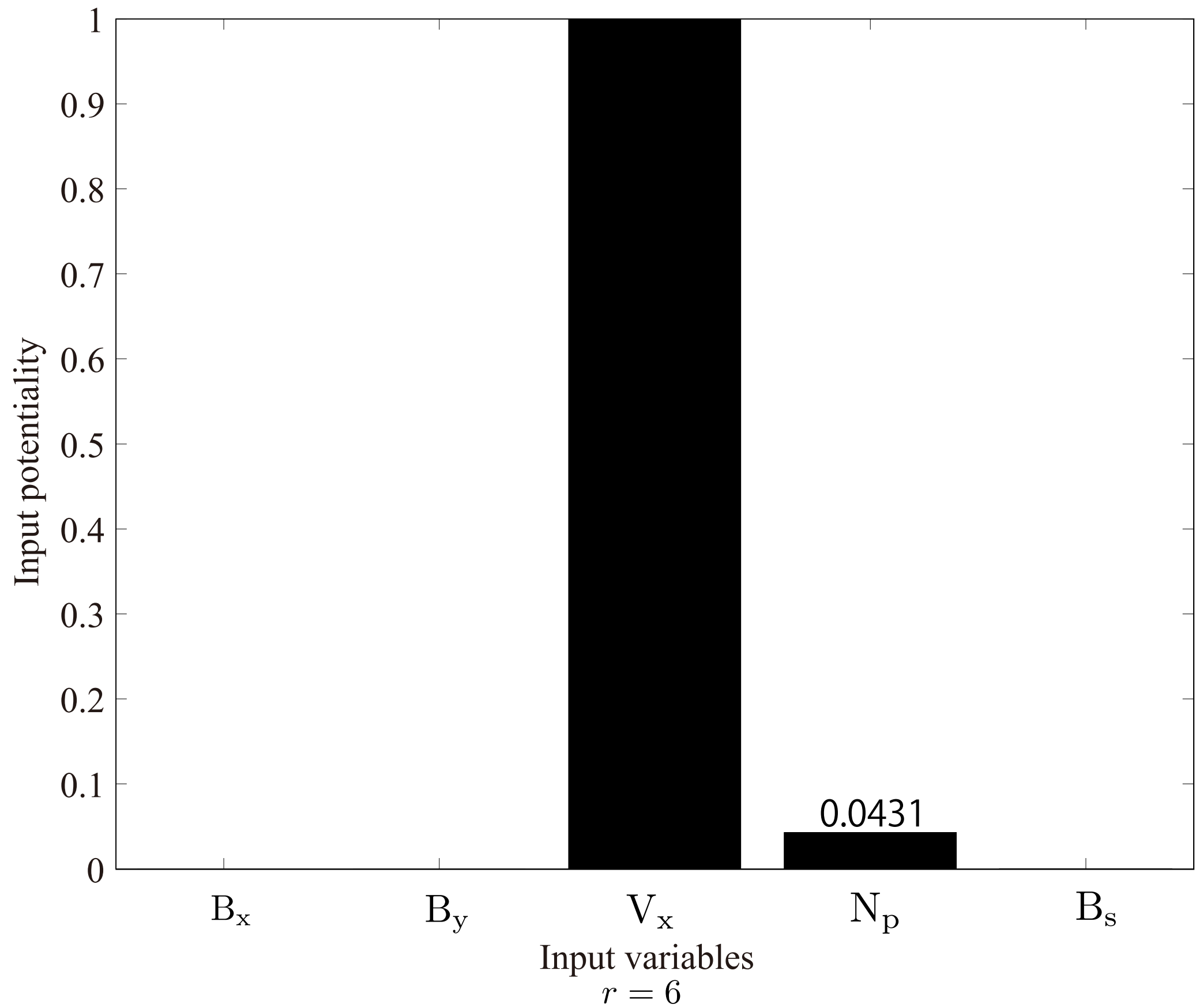

Figure 3

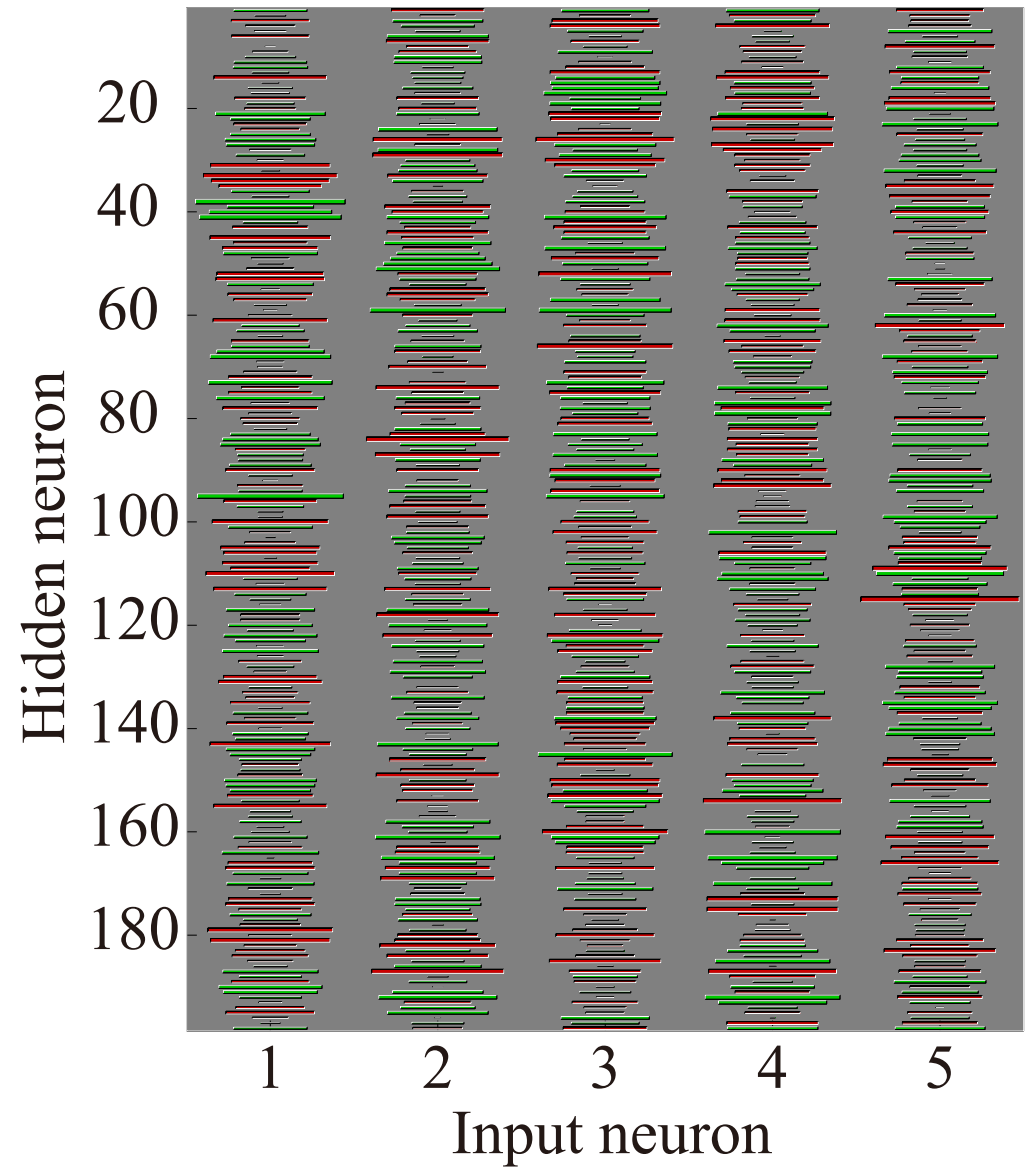 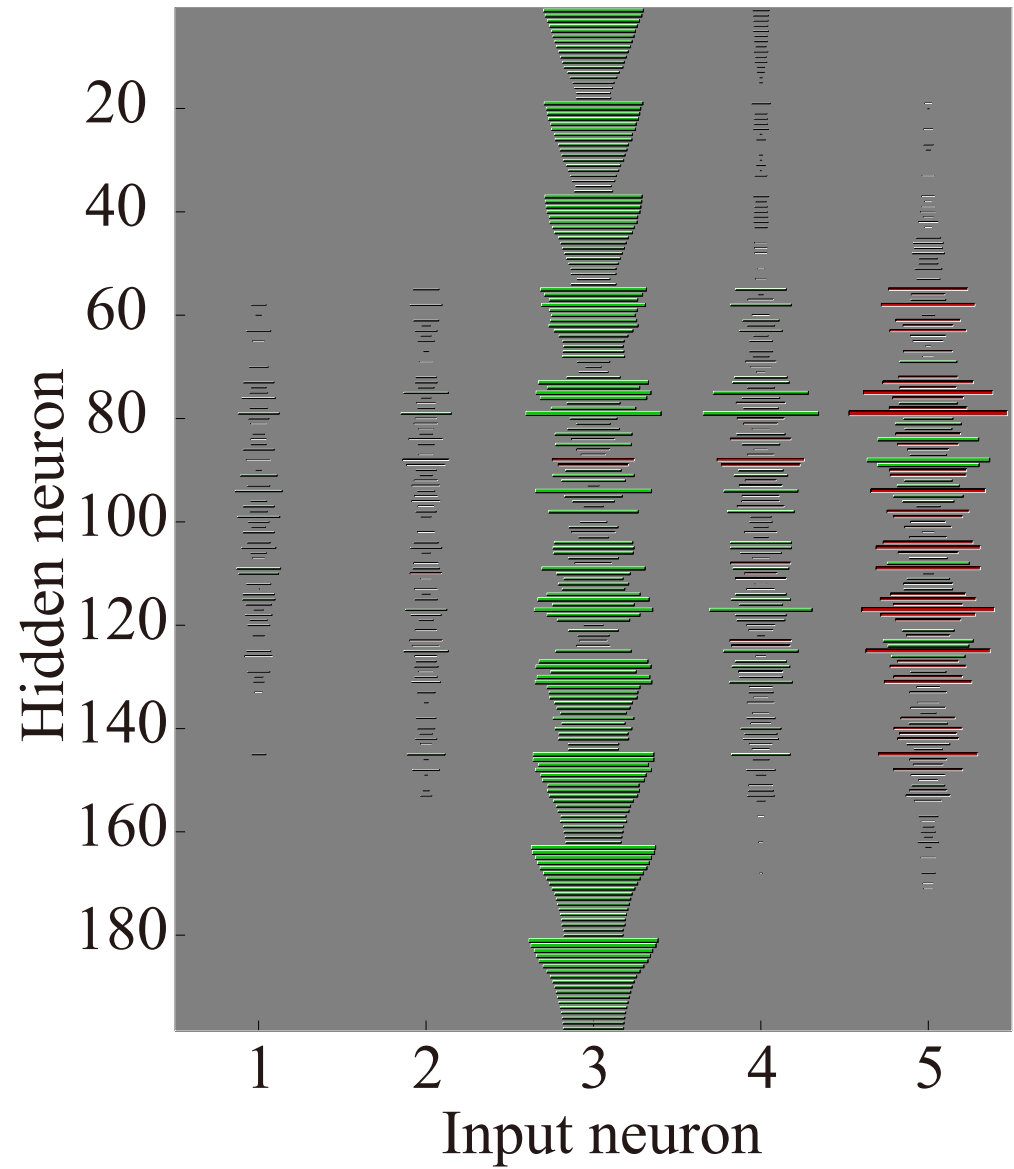

Figure 4

(a) MLP    (b) PL $(r = 6)$

Table 1

| No. | Input Parameters | Unit |
|---|---|---|
| 1 | $B_X$ [IMF GSE-X component] | nT |
| 2 | $B_Y$ [IMF GSE-Y component] | nT |
| 3 | $V_X$ [Solar wind velocity] | km/s |
| 4 | $N_P$ [Ion number density] | /cm$^3$ |
| 5 | $B_S^\dagger$ [Southward IMF-$B_z$] | nT |
| 6 | $K_P$ Index | - |

$$B_S^\dagger = \begin{cases} 0 & (B_z > 0), \\ B_z & (B_z \leq 0). \end{cases}$$

| Setup of PL | |
|---|---|
| # of input neurons | 5 |
| # of output neurons at Knowledge Accumulation step | 198 |
| # of hidden neurons at Knowledge Utilization step | 198 |
| # of output neurons at Knowledge Utilization step | 2 |

|  | PL | | | | | | | | | | MLP |
|---|---|---|---|---|---|---|---|---|---|---|---|
|  | $r=1$ | $r=2$ | $r=3$ | $r=4$ | $r=5$ | $r=6$ | $r=7$ | $r=8$ | $r=9$ | $r=10$ |  |
| Accuracy | 0.9828 | 0.9824 | 0.9824 | 0.9836 | 0.9824 | **0.9840** | 0.9819 | 0.9807 | 0.9828 | 0.9832 | **0.9903** |
| Precision | 0.9793 | 0.9793 | 0.9785 | 0.9801 | 0.9792 | 0.9801 | 0.9784 | 0.9768 | 0.9785 | **0.9817** | **0.9867** |
| Recall | 0.9866 | 0.9857 | 0.9866 | 0.9874 | 0.9857 | **0.9882** | 0.9857 | 0.9849 | 0.9874 | 0.9849 | **0.9941** |
| F-measure | 0.9828 | 0.9824 | 0.9824 | 0.9837 | 0.9824 | **0.9841** | 0.9820 | 0.9807 | 0.9828 | 0.9832 | **0.9904** |

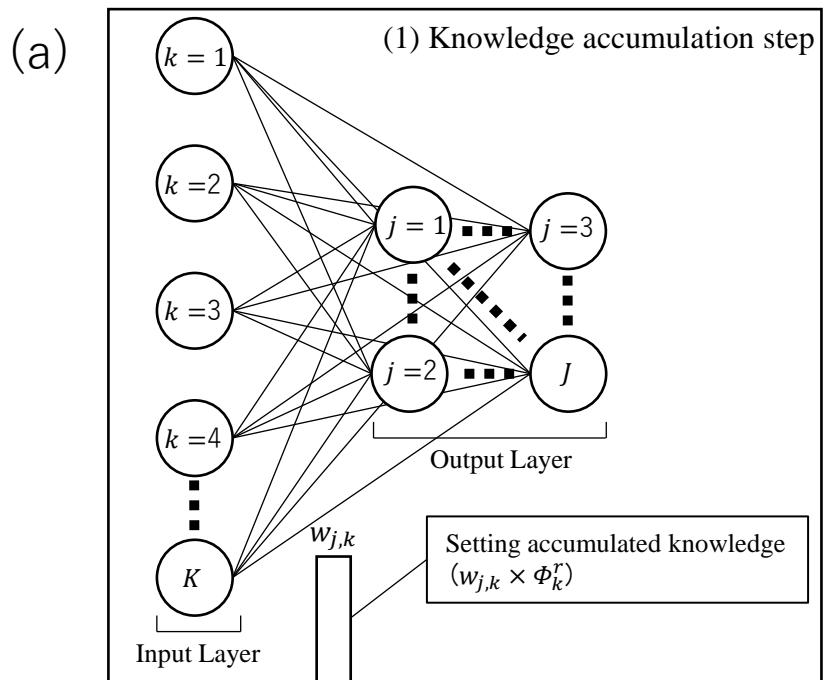

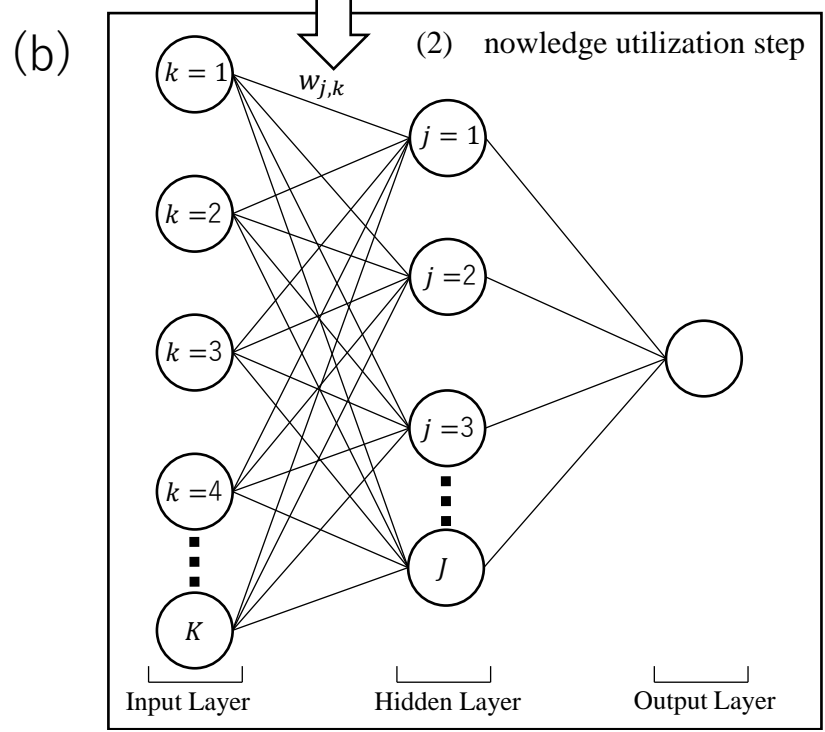

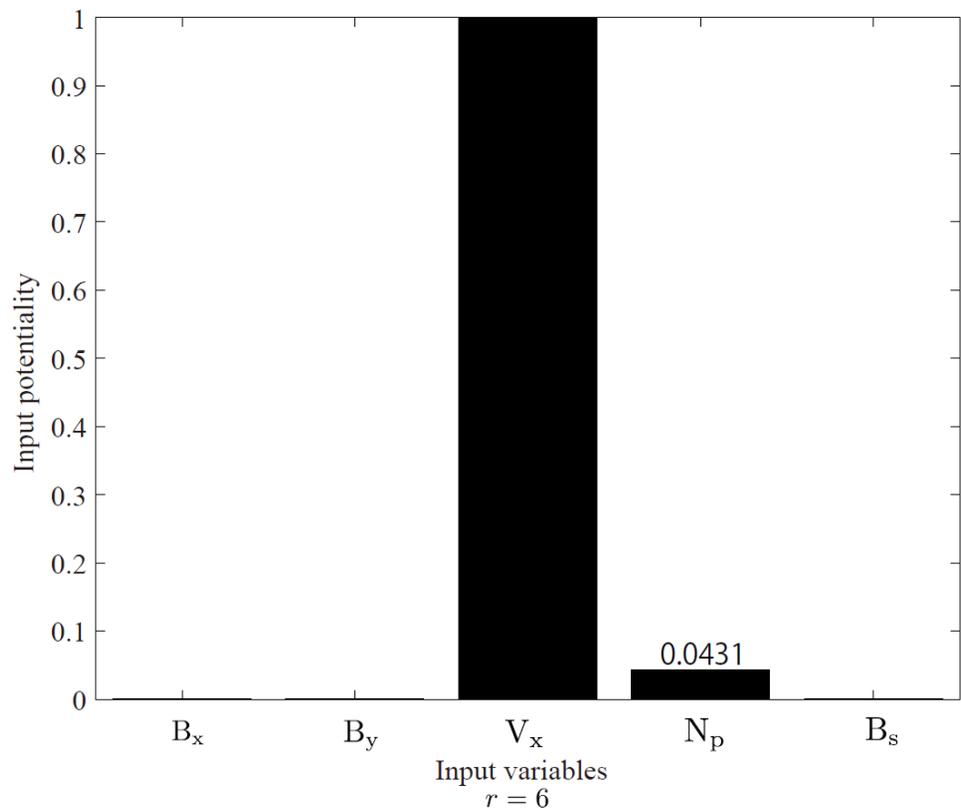

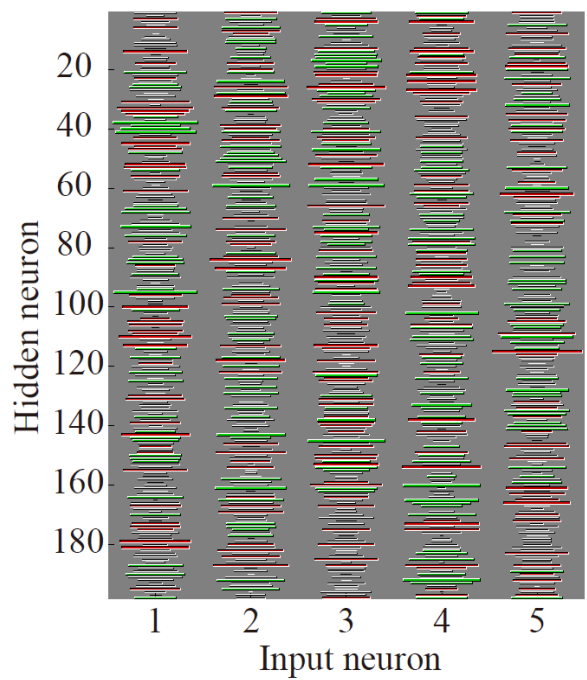 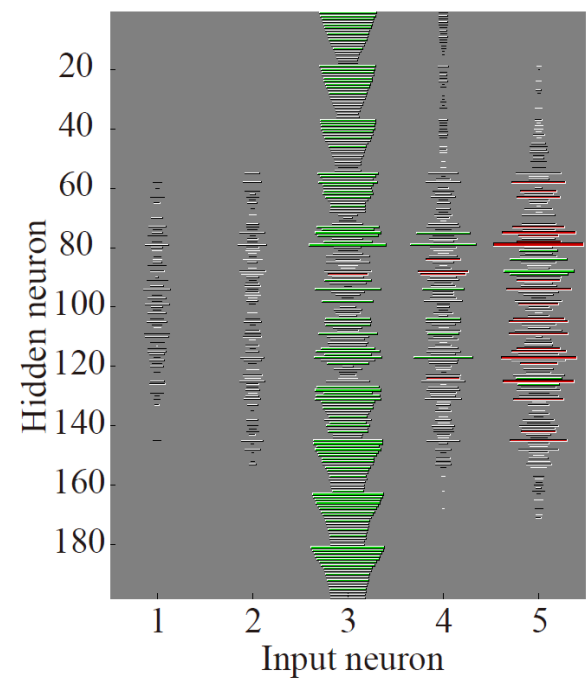